\documentclass[conference]{IEEEtran}
\usepackage{bm,amsfonts,amssymb,amsmath}
\usepackage[utf8]{inputenc} 
\usepackage{graphicx}

\begin{document}




\title{Exploring brain transcriptomic patterns: a topological analysis using spatial expression networks}  

\author{\IEEEauthorblockN{Zhana Kuncheva}
\IEEEauthorblockA{Department of Mathematics\\
Imperial College London, UK\\
Email: z.kuncheva12@imperial.ac.uk}
\and
\IEEEauthorblockN{Michelle L. Krishnan}
\IEEEauthorblockA{Perinatal Imaging and Health\\King's College London, UK\\
E-mail: michelle.krishnan@kcl.ac.uk}
\and
\IEEEauthorblockN{Giovanni Montana}
\IEEEauthorblockA{Biomedical Engineering Department\\King's College London, UK\\
E-mail: giovanni.montana@kcl.ac.uk}}
\maketitle
\begin{abstract}
Characterizing the transcriptome architecture of the human brain is
fundamental in gaining an understanding of brain function and disease. A
number of recent studies have investigated patterns of brain gene expression
obtained from an extensive anatomical coverage across the entire human brain using experimental data generated by the Allen
Human Brain Atlas (AHBA) project. 
In this paper, we propose a new representation of a gene's transcription
activity that explicitly captures the pattern of spatial co-expression
across different anatomical brain regions. For each gene, we define a
Spatial Expression Network (SEN), a network quantifying co-expression
patterns amongst several anatomical locations. Network similarity measures
are then employed to quantify the topological resemblance between pairs of
SENs and identify naturally occurring clusters. Using network-theoretical
measures, three large clusters have been detected featuring distinct
topological properties. We then evaluate whether topological diversity of
the SENs reflects significant differences in biological function through a
gene ontology analysis. We report on evidence suggesting that one of the
three SEN clusters consists of genes specifically involved in the nervous
system, including genes related to brain disorders, while the remaining two
clusters are representative of immunity, transcription and translation.
These findings are consistent with previous studies
showing that brain gene clusters are generally associated with one of these
three major biological processes.

\end{abstract}

\section{Introduction}

The human brain is a complex interconnected structure controlling all
elementary and high-level cognitive tasks~\cite{Oldham2008}. This complexity
is a result of the cellular diversity distributed
across hundreds of distinct brain anatomical structures~\cite%
{Hawrylycz2011,Hawrylycz2012}. One of the main tasks of the neuroscience
community in the past decade has been to connect the underlying genetic
information of the anatomical structures to their underlying biological
function~\cite{Hawrylycz2012,Hawrylycz2015,Richiardi2015}. 
A useful data source for such studies is the Allen Human Brain Atlas (AHBA)~\cite%
{Hawrylycz2012}{}, which provides microarray expression profiles of almost
every gene of the human genome with emphasis on an extensive anatomical coverage across the entire human brain.

In this paper, we make use of the experimental data provided by the AHBA
project to study the spatial microarray variability at the single gene
level. Analyzing the complete transcription architecture of the human brain
in this way may be informative of the impact of genetic disorders on
different brain regions that would otherwise not be
apparent due to the coarse resolution. 

To gain new insights into the expression patterns of the human brain and
identify potentially important biomarkers, many studies involving
the AHBA data explore gene to gene relationships~\cite%
{Hawrylycz2012,Hawrylycz2015}{}. Each gene is represented by its expression
levels across anatomical locations. Genes with correlated expression
profiles are grouped together based on an appropriate similarity measure. The analysis of the resulting gene co-expression networks provides evidence
that transcriptional regulation relates to anatomy and brain function~\cite%
{Hawrylycz2011,Hawrylycz2012,Hawrylycz2015}{}. There are also studies
that consider the genetic similarity between pairs of regions, and show that transcriptional
regulation varies enormously with anatomic location~\cite%
{Hawrylycz2012,Hawrylycz2015,Mahfouz2015,Goel2014}{}. These findings indicate
the necessity to adopt a new representation of a gene's transcription
activity that explicitly captures the pattern of spatial co-expression
across different anatomical brain regions. 

We propose a new and unexplored way to model the spatial variability at the single
gene level. For each gene, we create a spatial expression network, or SEN.
Each node of the network corresponds to a pre-defined brain region for which
we have sufficient transcriptomic data, and each edge weight represents the
similarity in gene expression levels, for that gene, between two brain
regions. Applying this procedure to genes that have been found to be stably
expressed across specimens gives rise to a population of approximately $17,000
$ gene networks, each one representing a brain-wide spatial pattern of gene
expression. Using this representation, we investigate whether the
topological similarity of the SENs reflects the biological similarity of
genes through an integrative analysis based on network clustering and gene
ontologies. Our hypothesis is that, if clusters of topologically similar
SENs can be identified, the corresponding genes within each cluster may also
share similar biological properties.

\begin{figure*}[t]
\centerline{
		\includegraphics[width=0.98\textwidth]{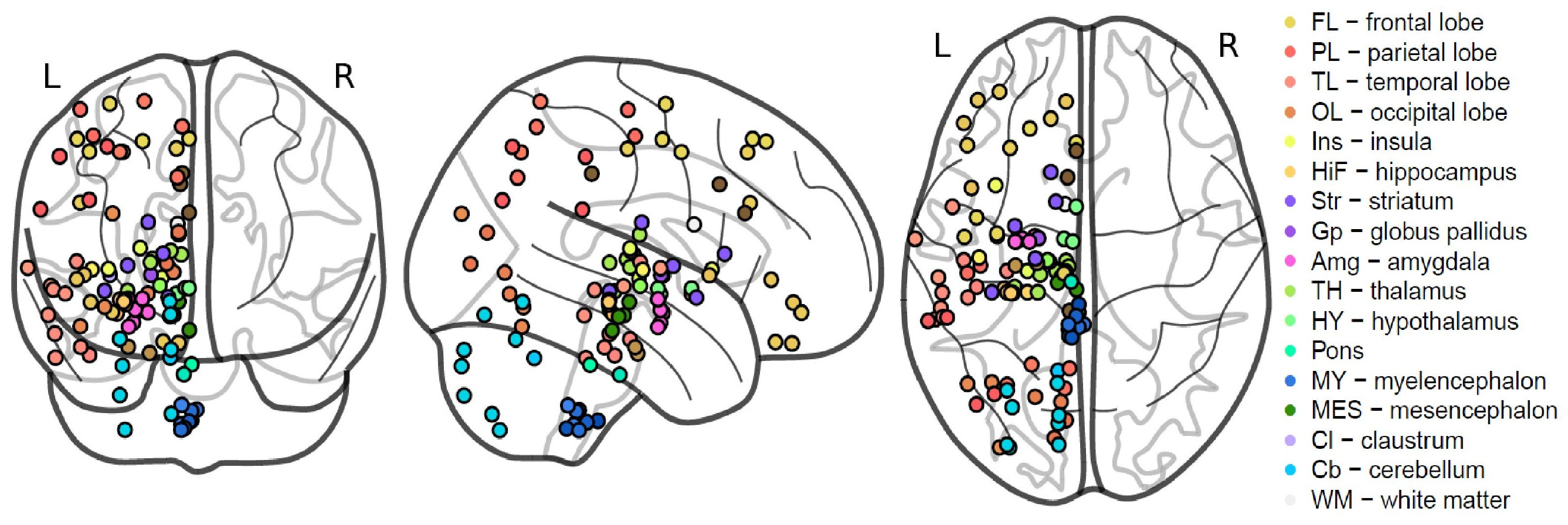}}  
\caption{Anatomical maps of the $105$ brain regions used to construct the
SENs. The maps show the brain regions as seen from inferior, lateral and
superior views, from left to right. All regions are in the
left hemisphere and they are located in the Thalamus, Cerebellum, Pons,
Midbrain, Medulla and Cerebral cortex. Coloring of the regions is consistent
with anatomical tissue and is obtained from AHBA ontology atlas \protect\cite%
{ahba2}.}
\label{fig:Figure1a}
\end{figure*}
A robust cluster analysis of all SENs has indicated the presence of three
large and stable clusters of SENs, each one having significantly different
topological features as well as different biological function. In
particular, one of the clusters has been found to be uniquely enriched for
brain-related terms, neurological diseases and genes with enriched
expression in neurons. Overall, our analysis provides evidence supporting
the notion that topological proximity of spatial gene networks is indicative
of similar biological function.


\section{Materials and Methods}

\subsection{Spatial Expression Networks (SENs)}

The Allen Human Brain Atlas (AHBA)~\cite{ahba2,Hawrylycz2012} is a publicly
available atlas of the human brain with microarray-based
genome-wide transcriptional profiling of specific brain regions spanning all
major anatomical structures of the adult brain. The data set includes
transcriptional profiling data from more than $3500$ samples comprising
approximately $200$ brain regions in $6$ clinically unremarkable adult human
brains. The Agilent $4\times 44$ Whole Human Genome platform was used for
gene expression extraction. Two donors contributed samples representing
approximately $1000$ structures across the whole brain, while the other four
approximately $500$ samples from the left hemisphere. Our analyses is based
on $16,906$ pre-selected genes from a previous study~\cite{Richiardi2015}{}.
We use the normalized expression levels, which were normalized across samples and across different
brains as in previous analyses~\cite{ahba}{}.

For each of the $16,906$ genes, we constructed an individual spatial
expression network (SEN) representing patterns of expression variability in
the brain. Only brain regions with at least one measurement in all $6$
brains were included in the analysis resulting in a total of $N=105$ regions
from the left hemisphere, as shown in Fig.~\ref{fig:Figure1a}.

The mean expression level for a gene in brain region $i$ is denoted by $g_{i}
$. The distribution of the mean and median values for each brain region over all genes were not found statistically different (Kolmogorov-Smirnov test~\cite{Marsaglia2003}{}; all $p>>0.05$). Furthermore, for more than $97\%$ of all region samples across all genes, the standard deviation of the expression values is less than $20\%$ of the mean value, indicating that the mean can be taken as representative of the expression values at a given region for a given gene.

Formally, we define a SEN as a fully connected network $G=(V,E)$ with node set $%
V=\{i:i=1,2,...,N\}$ indicating the brain regions and weighted edge set $%
E=\{E_{ij}:i,j=1,2,...,N;i\neq j\}$. Each edge weight $E_{ij} \in [0,1]$
quantifies the similarity in gene expression between regions $i$ and $j$.
The maximum value is reached when the mean expression levels in the two
brain regions are equal. We impose that $E_{ij}$ monotonically
decreases with an increasing absolute difference between mean expression
levels; accordingly, the edge weights are defined as 
\[
E_{ij}:= \frac{1}{1+|g_{i}-g_{j}|}. 
\]
This network representation allows us to capture the interconnected
variability of gene expression across the brain at the gene level.


\subsection{Clustering SENs \label{sec:LTO}}

In order to address our hypothesis that topological similarity may reflect
biological similarity, initially we set out to explore whether SENs form naturally
occurring clusters. For this we first required an appropriate measure of
topological dissimilarity between pairs of SENs. We first
mapped each SEN $G$ to a $N$-dimensional feature vector $\bm{d}%
=\left(d_{1},d_{2},...,d_{N}\right)$ with each elements representing the
node degree, i.e. $d_{i}=\sum_{j=1}^{N}E_{i,j}$. The degree for each node
captures the global transcriptomic similarity of the corresponding brain region
to all other brain regions for a given gene. If the node degrees for two SENs are very different,
then the corresponding genes have very different global transcriptomic patterns.
The dissimilarity between
two SENs, $G_l$ and $G_k$, was taken to be the Euclidean distance
between the corresponding feature vectors, $\bm{d}_l$ and $\bm{d}_k$.




Three different clustering algorithms were used -- partitioning around
medoids (PAM)~\cite{Kaufman1990}{}, k-means~\cite{VonLuxburg} and fuzzy 
\textit{C}-means~\cite{Dunn2008}{} -- all providing a partition of all the
SENs into $k$ different clusters. To determine an appropriate number of
clusters $k$ using each one of these algorithms we performed a stability
analysis~\cite{VonLuxburg}{}. The $k$ clusters are deemed ``stable'' if random changes
in the SEN configurations generate almost identical $k$ clusters. 
To introduce random changes in the networks, we use a
randomization strategy by which the observed networks in network space $\Gamma$ are perturbed slightly. For this analysis we used two different randomization procedures: (a) vertex permutations, i.e. we permuted the node
labels of a random subset of networks so as to preserve the node degrees but
not their order, (b) edge perturbation, i.e. we perturbed the edge weights of a
random subset of networks so as to make the cluster robust against white
noise. 

To obtain a measure of cluster instability, we use the following steps: First, we generate perturbed versions $\Gamma_{b}\left(
b=1,2,...,b_{\max}\right)$ of $\Gamma$, and cluster the networks in $\Gamma_{b}$ into $k$ clusters thus obtaining $\mathcal{C}_{b}(k)$. In addition, we randomize the cluster assignments~\cite{Lange} in $\mathcal{C}_{b}\left(  k\right)  $ to
obtain random clustering $\mathcal{C}_{b,\text{rand}}\left(  k\right)  .$
Second, for $b,b^{\prime}=1,2,...,b_{\max}$, we compute the pairwise distances
$\left[  1-NMI(\mathcal{C}_{b}(k),\mathcal{C}_{b^{\prime}}(k))\right]  $
between the clusterings $\mathcal{C}_{b}(k)$ and $\mathcal{C}_{b^{\prime}%
}(k),$ and between the randomized clusterings $\mathcal{C}_{b,\text{rand}%
}(k)$ and $\mathcal{C}_{b^{\prime},\text{rand}}(k)$. The normalized mutual information (NMI) is used as a
similarity measure between partitions~\cite{Meila2007}{}. The
cluster instability index is defined as the mean distance between clusterings $\mathcal{C}%
_{b}(k)$, i.e.

\begin{equation}
I(k)=\frac{1}{b_{\max}^{2}}\sum_{b,b^{\prime}=1}^{b_{\max}}\left[
1-NMI(\mathcal{C}_{b}(k),\mathcal{C}_{b^{\prime}}(k))\right]  .\label{eq:1}%
\end{equation}
We use the normalized instability  index, $I_{\text{norm}}(k):=I(k)/I_{\text{rand}}(k)$, which corrects for a scaling~\cite{Lange}
of $I(k)$ with an increasing number of clusters~$k$. We choose number of
clusters $k$ that gives the lowest $I_{\text{norm}%
}(k).$ 



\subsection{Topological characterization of SEN clusters\label{sec:network}}

To characterize the topological properties of SENs in each cluster, we use global topological measures that capture different aspect of the network such as its density, the tendency of its nodes to cluster and form communities, the presence of central and hub nodes. Overall, we use eight such different measures: average node degree~\cite%
{Opsahl2010}{}, average closeness centrality~\cite{Opsahl2010}{}, weighted
diameter~\cite{Barrat2004}{}, global clustering coefficient for weighted
networks~\cite{Barrat2004}{}, number of non-overlapping communities, average
authority score~\cite{Kleinberg1999}{}, the number of nodes with authority
score $>0.95$, and the number of nodes with authority score $<0.05$. All
measures were computed for all SENs within each cluster. To test for
statistically significant differences in network topology across clusters,
we performed a multivariate ANOVA test~\cite{Scheiner2001}{}.

Furthermore, for each SEN we derived a measure of community structure~\cite%
{Fortunato2010}{}. In our context, the presence of a community in a given SEN
indicates that there is a set of highly interconnected brain regions whose
gene expression similarity is higher compared to the rest of the network.
For this analysis we used the Fast Greedy algorithm~\cite{Clauset2004}{},
which is based on the optimization of the modularity function that sums the
edge weights within a community and corrects for the expected edge weights
by chance. The algorithm is discriminative of small edge weight differences
and can yield sensitive separation of brain regions into communities. Genes
with similar community structures indicate the presence of similar local
coherent transcriptomic patterns for groups of brain regions.

For each cluster, we quantify the similarity of a pair of brain regions
using the communities detected in all the SENs by counting the number of
times the two regions fall within the same community. This count is then
divided by the total number of SENs in the cluster in order to obtain an
index lying in the $[0,1]$ range, which we call the ``coherence index''.
Values close to $1$ indicate high coherency between the two brain regions,
i.e. the average tendency to fall within communities of highly
interconnected brain regions.

\subsection{Biological characterization of SEN clusters\label{sec:enrichment}%
}

In order to investigate whether naturally occurring clusters formed by SENs
can be related to distinct biological function, we require a procedure which assigns representative biological terms to each cluster.
For this purpose we use a Gene Ontology (GO) enrichment analysis pipeline
which first collects broad GO information for the biological context
of genes in each of the main clusters, and then reduces this information to
representative GO terms for final interpretation of the clusters.

Each SEN cluster was first annotated for significantly enriched Biological
Process (BP) terms using a standard hypergeometric test for over-represented
terms ($p<0.001$) implemented in the GOstats R package~\cite{Falcon2007}{}.
Using a clustering methodology implemented in the tool REVIGO \cite{Supek2011}{}, we group semantically similar GO terms based on the established 
\textit{SimRel} measure. The algorithm finds a representative term for each
group based on the enrichment p-values, with a bias away from very general
parent GO terms. The size of the resulting summary list is controlled by
setting the threshold for the~\textit{SimRel} similarity measure at $0.5$.
Results are summarized by retaining the cluster representatives for each GO
term that can reveal underlying function of these clusters.

Genes in each of the clusters were also annotated for disease enrichment
using the WebGestalt tool~\cite{Zhang2005a}{}, which interfaces with the GLAD$4
$U platform~\cite{Jourquin2012} to retrieve and prioritize disease-gene
links from publications, using a hypergeometric test with multiple testing
correction and the genome as background.

\section{Experimental results}

\subsection{Topologically different SEN clusters\label{sec:res}}

\begin{figure*}[!t]
\centerline{
\includegraphics[scale=0.5]{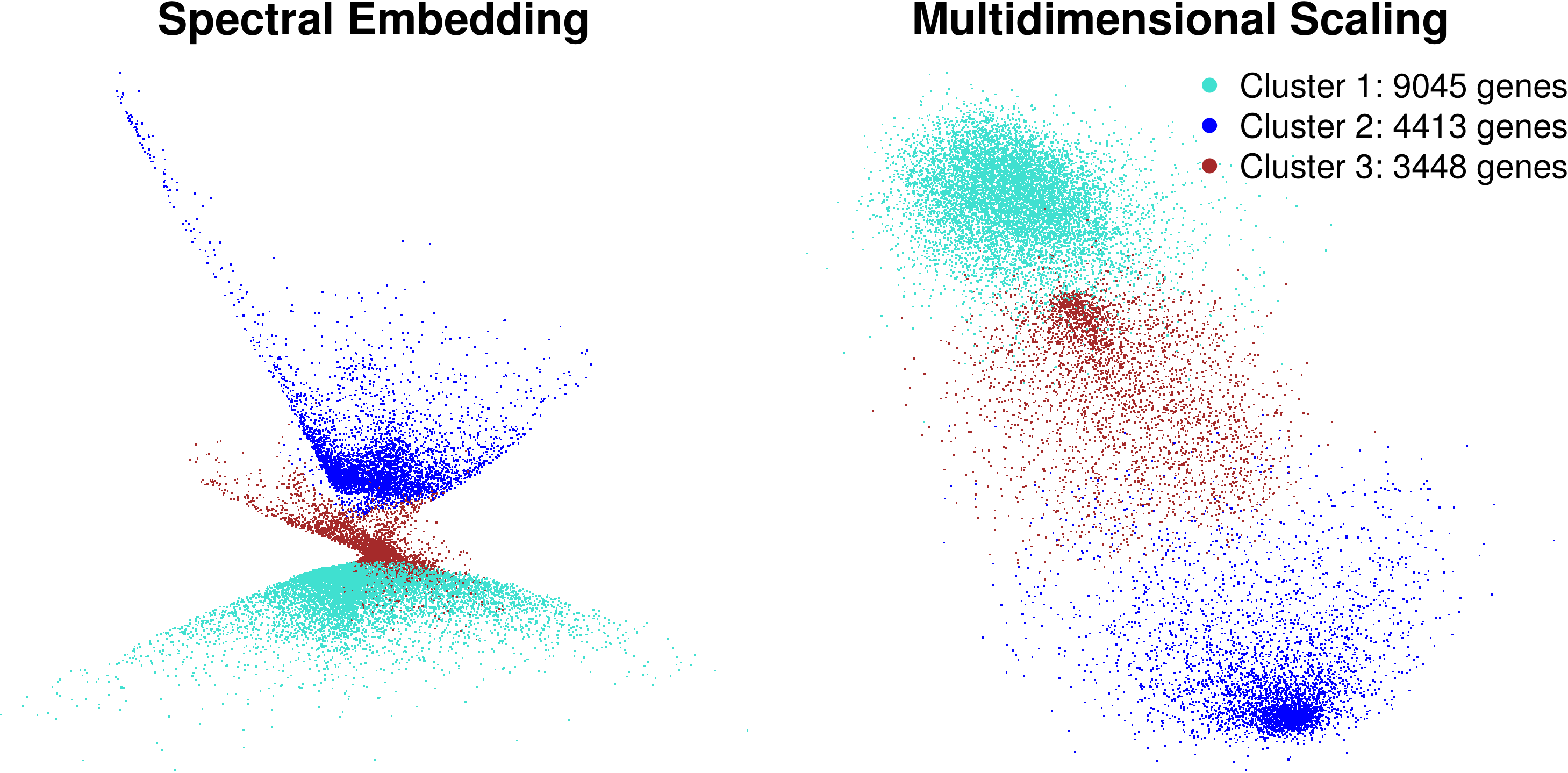}}
\caption{Two-dimensional visualization of all SENs using two different dimensionality
reduction algorithms: spectral embedding \protect\cite{Luxburg} (left) and
multidimensional scaling \protect\cite{Borg2005} (right). The color scheme
indicates the cluster membership as determined by the PAM algorithm. Both visualizations indicate three main clusters.}
\label{fig:Figure2}
\end{figure*}

All SENs were clustered into up to six clusters using the procedures
outlined in Sec.~\ref{sec:LTO}. The two instability analyses were each
performed using $b_{\max}=500$. Using the first randomization scheme, $5\%$
of networks were randomly sampled for node permutation, while in the second
procedure $20\%$ of networks were randomly sampled and white noise was
introduced by adding $\pm20\%$ to each edge weight. The results for all
three clustering procedures, Tab.~\ref{tab:table1}, show that PAM clustering
has the lowest instability followed by fuzzy \textit{C}-means. Furthermore,
for all three clustering methods grouping data into two and three clusters
leads to the lowest instabilities. 

\begin{table}[h]
\centering
\caption{ Different stability analyses for three different clustering algorithms using two
randomization strategies (vertex and edge permutation).} 
{
\begin{tabular}{lrrr|rrr}
\hline
 & \multicolumn{3}{c|}{Vertex permutation} & \multicolumn{3}{c}{Edge
perturbation} \\ \hline
 $I_{\text{norm}}(k)$ & PAM & Cmeans & k-means & PAM & Cmeans & 
k-means \\ 
 $I_{\text{norm}}(2)$ & 0.016 & 0.020 & 0.065 & 0.009 & 0.015 & 0.065
\\ 
$I_{\text{norm}}(3)$ & \textbf{0.018 }& \textbf{0.023 }& \textbf{0.076} & \textbf{0.010 }& \textbf{0.016} & \textbf{0.071 }\\ 
$I_{\text{norm}}(4)$ & 0.023 & 0.031 & 0.092 & 0.019 & 0.033 & 0.180 \\ 
$I_{\text{norm}}(5)$ & 0.026 & 0.038 & 0.171 & 0.025 & 0.030 & 0.191 \\ 
$I_{\text{norm}}(6)$ & 0.031 & 0.080 & 0.187 & 0.027 & 0.086 & 0.208 \\ \hline

\end{tabular}%
}
\label{tab:table1}
\end{table}

The PAM algorithm was chosen to generate the final partitions as it yields the
lowest instability index. As an additional validation to support the choice
of three PAM clusters, we used three internal validation measures: the
Sillhouette width~\cite{Rousseeuw1987}{}, the Dunn index~\cite{Dunn2008} and
the within-cluster variance~\cite{Halkidi2001}{}. The Dunn index and
Silhouette width support the presence of two to three clusters, see Tab.~\ref%
{tab:table2}. However, the intra-cluster variance, which is known to be more
sensitive to the existence of sub-clusters~\cite{Liu2010}{}, shows that
grouping data into two clusters leads to high within-cluster variability
compared to a higher number of clusters. By taking all these criteria into
account, we have chosen to consider $k=3$ since this leads to the 
lowest instability and within-cluster variability whilst having as high as possible Dunn and Sillhouette scores. 
\begin{table}[h]
\centering
\caption{ Cluster validation measures for clustering SENs into $k$ clusters using
PAM.} {
\begin{tabular}{lrrr}
\hline
 \textit{k} & Dunn & Silhouette & Within-cluster Variance \\ \hline
 2 & 2.20 & 0.66 & 0.276 \\ 
3 & \textbf{1.20} &\textbf{ 0.44} & \textbf{0.225} \\ 
4 & 0.61 & 0.30 & 0.215 \\ 
5 & 0.63 & 0.23 & 0.223 \\ 
6 & 0.48 & 0.18 & 0.211 \\ \hline

\end{tabular}
} 
\label{tab:table2}
\end{table}

In an attempt to visually assess whether this choice seems appropriate, we
used a distance-preserving projection of all $16906$ SENs into a $2$D-dimensional space using two different dimensionality reduction procedures:
spectral embedding~\cite{Luxburg} and multidimensional clustering~\cite%
{Borg2005}{}. The resulting projections can be found in Fig.~\ref{fig:Figure2}%
. All three clusters -- $1$ (turquoise), $2$ (blue) and $3$ (brown) --
appear well-separated. 

\begin{figure*}[t]
\centerline{
		\includegraphics[width=0.95\textwidth]{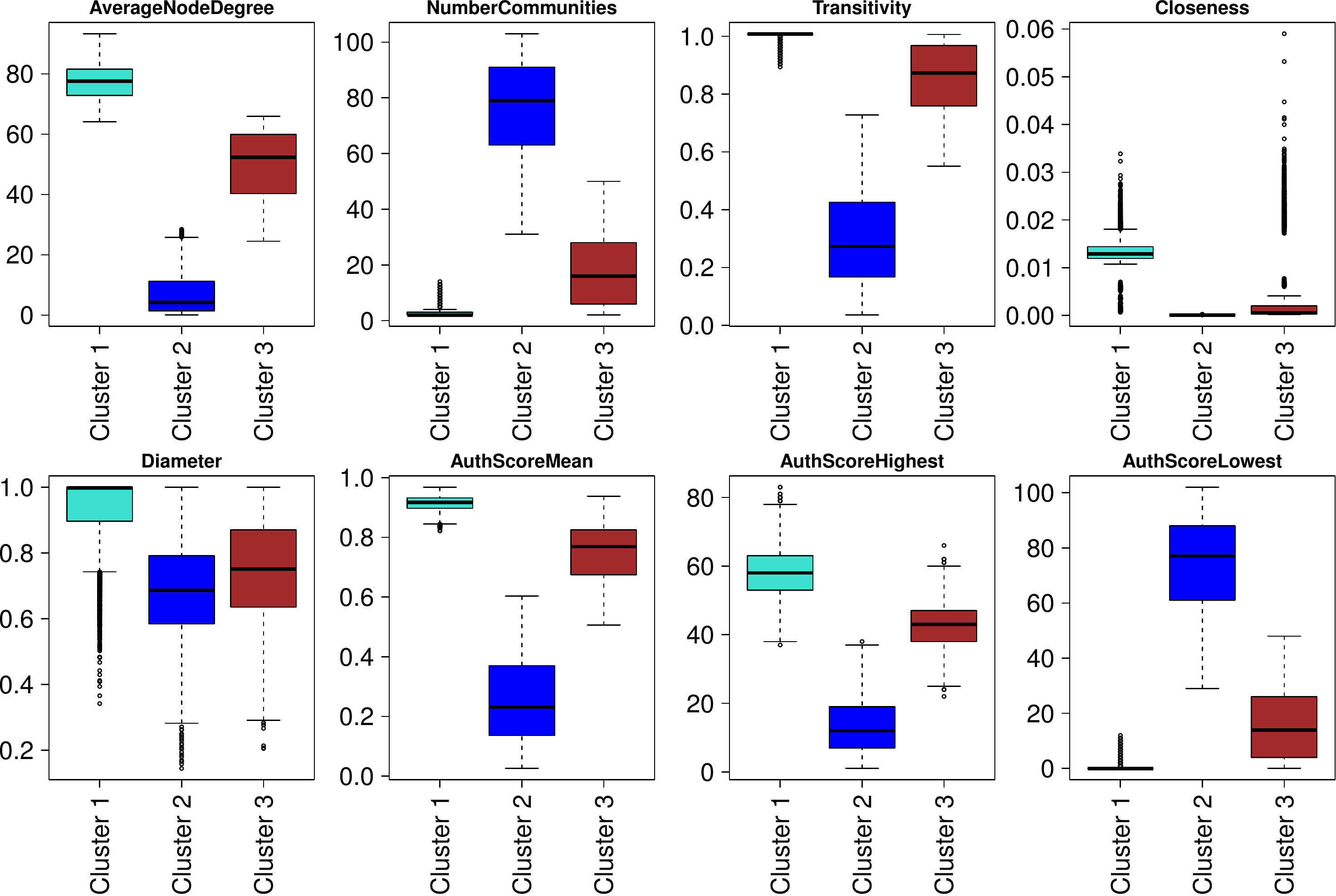}}  
\caption{ Distribution of topological network measures in the three clusters
obtained using the PAM algorithm. High node degrees imply high edge weights with fewer low-weighted shortest
paths and fewer discrepancies in edge values. This leads to high
transitivity and closeness values, simultaneously reducing the number of
communities SENs are partitioned into. Higher node degrees lead to more
nodes having high authority scores thus increasing both the average
authority scores and the number of nodes with high authority. Low node
degrees signify sparseness of the SENs and more low-weighted shortest paths.
This results in more nodes being grouped in their own communities, in
addition to low closeness and transitivity. Sparse networks and low node
degrees result in lower authority scores and fewer nodes with high authority
score.}
\label{fig:Figure3}
\end{figure*}

\subsection{Topological differences amongst SEN clusters\label{sec:top}}

To validate that the three SEN clusters have distinct topological structure,
we used the eight global network measures outlined in Sec.~\ref{sec:network}. The frequency distribution of the
topological measures for each cluster is summarized in Fig.~\ref{fig:Figure3}
where a clear mean difference can be observed for each individual measure
across clusters. Using a MANOVA test, we reject the null hypothesis
of equality of topological features across clusters ($p<2.2e-16$; Wilk's $%
\Lambda=0.3589$). 

We have found that Cluster $1$ mostly consists of SENs with the highest node
degree, centrality measures, diameter, authority score and number of nodes
with high authority score, while there are only a few number of communities
and few nodes with low authority score. These properties imply coherent
expression levels across all brain regions. On the other hand, Cluster $2$
comprises of SENs with the lowest node degree, centrality measures,
diameter, authority score and number of nodes with high authority scores, and the highest number of communities and
nodes with low authority score. This indicates that most SENs within this
cluster are sparse, and that there is high variability between expression
levels across brain regions. Finally, Cluster $3$ consists of SENs with
medium ranged values for all network measures, implying moderate
variability between expression levels across brain regions.

\subsection{Biological differences amongst SEN clusters\label{sec:bio}}

\begin{figure*}[!t]
\centering
\includegraphics[width=0.95\textwidth]{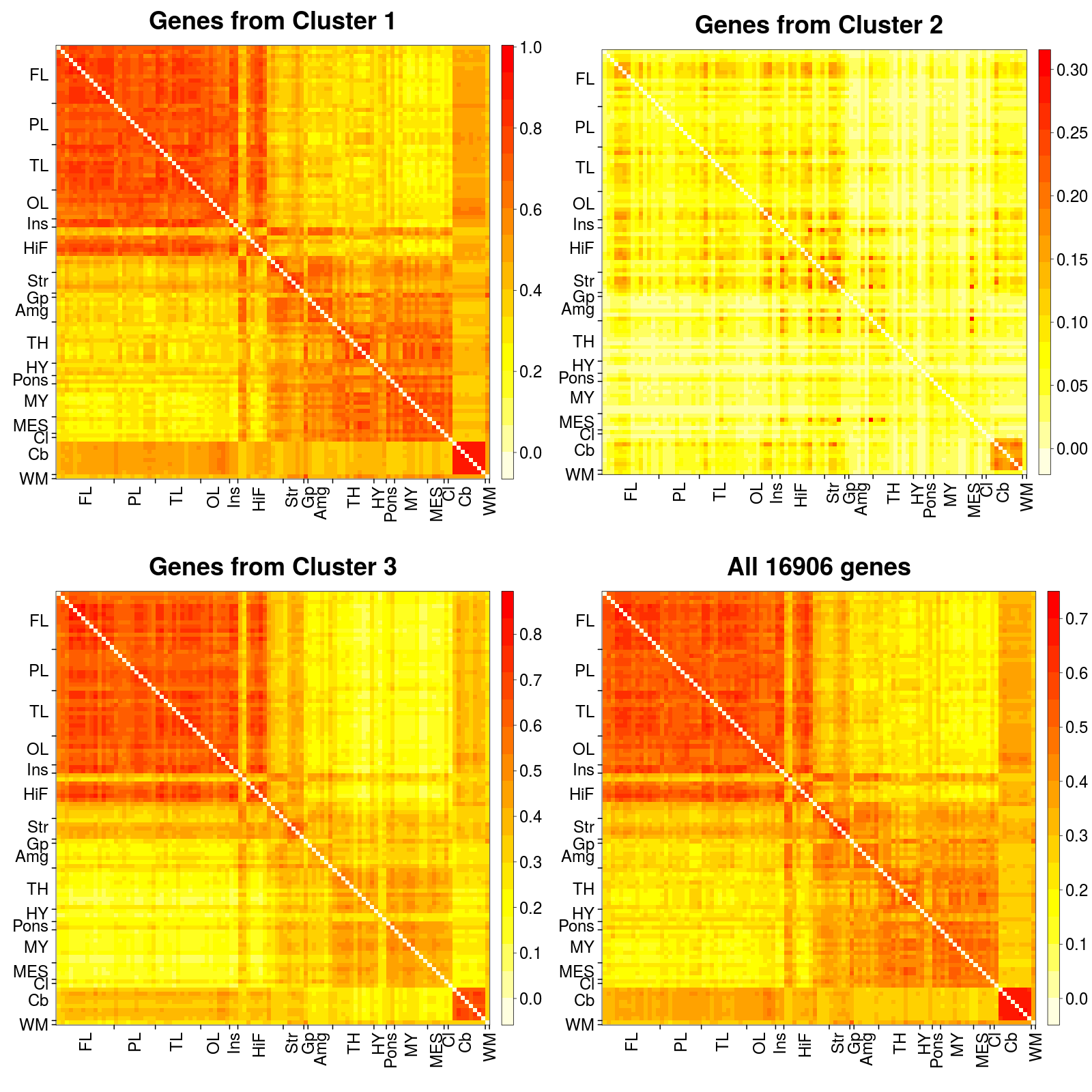}
\caption{Heatmaps representing the ``coherence index'' between pairs of brain
regions in each of the three SEN clusters and across all $16906$ genes. Each
pixel on the heatmap is the ``coherence index'' between the two corresponding
brain regions. Each heatmap is accompanied by a color key, where higher
values indicate high homogeneity of expression levels and lower values
indicate heterogeneous expression levels. The $105$ brain regions are mapped
to $17$ major brain structures using the AHBA ontology atlas \cite%
{ahba2} and abbreviated as indicated in Fig.~\ref{fig:Figure1a}.}
\label{fig:heatmap}
\end{figure*}

We investigated the local transcriptomic patterns within each of the three
clusters using the ``coherence index'' defined in Sec.~\ref{sec:network}. The
three clusters have different transcriptomic patterns, Fig.~\ref{fig:heatmap}%
, and comparing heatmaps of the three clusters to one for all $16906$ genes
shows that Cluster $1$ is closest to the genome-wide global patterning,
while Cluster $2$ and Cluster $3$ are carriers of imposed heterogeneity. The
patterns of the $16906$ genes are also consistent with existing work, and
largely replicate previous findings~\cite{Hawrylycz2012,Hawrylycz2015,Mahfouz2015}{}. In particular, homogeneity within
the Neocortex and Cerebellum, and increased heterogeneity in the Basal
Ganglia, have been previously reported. Cluster $2$ has few coherency
patterns in the Basal Ganglia regions and Cerebellum. Cluster $1$ exhibits
high homogeneity within the Cerebellum and the Neocortex, and between
subdivisions of the subcortical structure and the Hippocampus. Cluster $3$
appears to have coherent patterns in the Cerebellum and the Neocortex but
increased variability in the Basal Ganglia.

Obtaining detailed annotation as described in Sec.~\ref{sec:enrichment}
revealed that all three clusters are significantly enriched ($p<0.001$) for a variety
of GO BP terms. We reduced these large sets of GO terms to smaller
non-redundant sets by applying REVIGO~\cite{Supek2011}{}. 

The BP representative terms selected on the basis of enrichment $p$-values and
semantic similarity indicate that Cluster $1$ genes can be
described primarily by ``RNA processing'' and ``ribonucleoprotein complex
biogenesis''. Cluster $2$ genes are predominantly involved in immunity
including ``immune system process'', ``leukocyte proliferation'' and ``G-protein
coupled receptor signaling pathway'' terms primarily associated to the immune
system, whereas Cluster $3$ genes are uniquely involved in
``behavior'', ``metal ion transport'' and ``nervous system development''. On
closer inspection of Cluster $3$, these representative terms comprise
several linked biological processes specific to the Nervous System, and
which are not found on either Cluster $1$ or $2$, such as ``synaptic
transmission'' and ``dendrite extension''.

The significant disease enrichment (adjusted $p<0.001$) also supported the
functional distinctiveness of the three clusters, with Cluster $1$
being enriched for Mitochondrial disease, Cluster $2$ being significantly
enriched for genes involved with Immune System and Inflammatory disease, and
Cluster $3$ being principally involved in Nervous system disorders. Given
the observed functional differentiation between clusters, we investigated
whether this might correspond to cell-type specialization. We obtained lists
of neuron- and microglia-enriched genes in a repository of detailed
RNA-sequencing and splicing data from purified cell cultures~\cite{Zhang2014}{}
, and computed significant intersections using the SuperExactTest~\cite{Wang2015a}{}. This showed that genes in Cluster $3$ have significant
overlap with neuron- and microglia-specific genes ($p<0.05$). Cluster $2$,
on the other hand, has a unique association to microglia-specific genes only
($p<0.05$).


\section{Discussion}

Analyzing the transcriptome architecture of the human brain is a challenging
task due to the high-dimensionality and biological complexity of the data.
This is compounded by technical factors related to sample acquisition and
measurement error that can influence the results. We addressed the issue of
anatomical variability in gene expression by proposing to model each gene's
spatial co-expression pattern across anatomical regions as an individual
spatial network, or SEN. To explore whether topological similarity of gene
expression as captured by SENs is related to biological similarity, we used
network dissimilarity to obtain clusters of genes with similar patterns of
spatial co-expression. We aimed to gain additional insights into the
biological interpretation of regional anatomical specialization of the brain.

We demonstrated that there is evidence to support the presence of three
topologically distinct clusters of SENs, with each cluster being
characterised by particular network properties. 
Furthermore, investigating the community structure of the SENs, we identified possible
anatomical basis for the difference in the topological properties in the
three clusters. The differences between clusters are mainly due to the
heterogeneity of expression levels in the Basal Ganglia, and between the
Neocortex and Cerebellum.

We also found these three topologically distinct clusters to have
biologically distinct properties. On closer inspection we find Cluster $3$
to be specific to the nervous system, while Cluster $2$ appears to be
involved with immunity and Cluster $1$ with transcription and translation.
These associations are in line with previous results on the AHBA data set~\cite{Hawrylycz2012,Hawrylycz2015}{}, where the majority of clusters obtained
using WGCNA~\cite{Horvath2011}{}, a well-known gene clustering procedure, were also associated to immunity, nervous
system or transcription and translation. 

To gain an insight into possible cellular contributions to these
differences, we included cell-type specific data and observe that the
overlap of neuron- and microglia- specific genes in Cluster $3$ is in
keeping with current hypotheses regarding the significant interactions
between these two cell-types, including the possible modulatory activity of
microglia in synaptic pruning and cell communication beyond purely immune
functions~\cite{Tremblay2010}{}. 

We found significant disease associations for all three clusters, implying
the high biological impact of the genes involved and the utility of our
modular clustering approach for the identification of therapeutic targets.
There is a preponderance of neurological and neuropsychiatric conditions
linked to Cluster $3$ genes, and immune disorders linked to Cluster $2$,
reflecting their biological functions as described above and supporting
those annotations.

One important concern was whether the above results were specific to using
node degrees or they could be reproduced using other feature vectors. Thus
we constructed two different sets of feature vectors based on node
centrality as captured by the authority score and based on the raw edges of
the SEN. Based on each new set of feature vectors, results not included in this paper
demonstrated evidence to support the presence of three topologically distinct clusters of SENs. For both feature vectors, the three clusters were again
marked by different topological properties although there were shifts in the
distributions of those properties. Even so, in both cases the three clusters
were uniquely associated to the immune system, nervous system or
transcription and translation. 

For comparison purposes, we used WGCNA on the gene expression values of the $16906$ genes
for the $105$ brain regions. Results not included in this paper showed that WGCNA did not assign a cluster membership
to the majority of genes in Cluster $2$ due to the sparseness of their expression levels. More and smaller clusters were discovered with higher instability. 
The advantage of our method compared to WGCNA is that the structure of SENs allows us to use a number of
clustering procedures to detect stable gene clusters, whose validation could be achieved using
both topological and biological measures. We determine 
the biological function of a cluster using the gene ontology of the entire set of genes in the cluster, which is robust to slight changes in the cluster membership. 

A next step in the analysis of SENs should consider additional clusters to detect more specialized biological functions. Furthermore, it is well known that gene expressions in the cerebellum, subcortical and cortical regions differ significantly from each other based on their composition of different cell types~\cite{Hawrylycz2012,Hawrylycz2015}{}. Future work in this direction will include an analysis where only neocortex regions are used to construct SENs.

\section{Conclusion}
An important and challenging task in studying the brain transcriptional
architecture is integrating and modelling the high dimensionality of the gene
expression across the brain. To the best of our knowledge, our work is the
first to perform a region-wise comprehensive profiling of gene-specific
co-expression patterns across the human brain. By modelling gene expression
as SENs and employing network embeddings, we identified distinct clusters of
genes associated to specific biological functions, topological properties
and cell-types, with potential implications for neuropsychiatric disease.
Modelling genes as SENs across brain regions could be used for future studies
in helping to identify genes with particular co-expression patterns across a
set of spatial brain locations of interest, enabling the identification of
genes that act in spatially contextualized clusters with high biological
impact. As more microarray gene expression data become available at higher
spatial resolution and cell-type specificity, modelling gene co-expression
across the brain will be increasingly important to understanding the brain
transcriptome architecture at a microstructural scale.
\IEEEtriggeratref{17}
\bibliographystyle{hieeetr}
\bibliography{PaperLib}

\end{document}